\documentclass[12pt,letterpaper]{article}
\usepackage[latin1]{inputenc}
\usepackage{amsmath}
\usepackage{amsfonts}
\usepackage{amssymb}
\usepackage{mathtools}
\usepackage{graphicx}
\usepackage{geometry}
\usepackage{subcaption}
\usepackage{float}
\usepackage{wrapfig}
\usepackage{cite}
\title{A general model for thermal and electrical conductivity of binary metallic systems}
\author{Changdong Wei, Nikolas Antolin, Oscar D. Restrepo,\\ Wolfgang Windl*, and Ji-Cheng Zhao*}
\date{}
\begin{document}
	\maketitle
	\textit{Department of Materials Science and Engineering, The Ohio State University, Columbus, Ohio 43210}
	\break
	*Corresponding authors: windl.1@osu.edu (Wolfgang Windl) and zhao.199@osu.edu (J.-C. Zhao)
	
	\section*{Abstract}
	We extended and updated Mott{\textquoteright}s two-band model for the composition-dependence of thermal and electrical conductivity in binary metal alloys based on high-throughput time-domain thermoreflectance (TDTR) measurements on diffusion multiples and scatterer-density calculations from first principles. Examining Au-Cu, Au-Ag, Pd-Ag, Pd-Cu, Pd-Pt, Pt-Rh, and Ni-Rh binary alloys, we found that the nature of the two dominant scatterer-bands considered in the Mott model changes with the alloys, and should be interpreted as a combination of the dominant element-specific \textit{s}- and/or \textit{d}-orbitals. Using calculated orbital and element-resolved density-of-states values calculated with density functional theory as input, we determined the correct orbital mix that dominates electron scattering for all examined alloys and found excellent agreement between fitted models and experiments. This general model of the composition dependence of the thermal and electrical resistivity can be readily implemented into the CALPHAD framework.
	\bigbreak
	\textbf{Keywords:} Thermal conductivity; Electrical conductivity; Mott$^{+}$ two-band conductivity model; CALPHAD; First-principles; Density of states.
	
	\section{Introduction}
	Thermal conductivity is one of the most important thermophysical properties in engineering and scientific research, with applications in areas such as microelectronics \cite{ref14}, gas turbine engines \cite{ref15}, and thermoelectric devices \cite{ref16}. For instance, high thermal conductivity materials are needed in microelectronics to dissipate the accumulated heat generated by electronic components, while low thermal conductivity materials such as thermal barrier coatings (TBC) \cite{ref17} are required in jet engines to limit the exposure of key structural components to high temperature.  Thus, thermal conductivity data are critically needed in engineering product design and computational modeling of system performance \cite{ref18}.
	\bigbreak
	In metals and alloys, free electrons are the main carriers for both thermal and electrical conductions, which make the thermal and electrical conductivities proportionally related by the Wiedemann-Franz relation. There, phonon-induced lattice thermal conductivity only contributes less than 5\% to the total thermal conductivity \cite{ref19}. In alloys, solute atoms result in lattice perturbation and change the electron distribution, and thus significantly alter the electrical conduction behavior of the host materials. As a consequence, residual resistivity (which is the resistivity caused by electron interaction with the defects in the material) is mainly a function of alloy and defect concentration, and is therefore nearly independent of temperature.
	\bigbreak
	Based on these observations, conductivity models have been proposed that model the composition dependence of resistivity in metal alloys based on simple model assumptions, most notably by Nordheim \cite{ref20} and Mott \cite{ref21,ref22}. However, as we will show, the model assumptions underlying these models do not generally hold for all alloy systems, and not all systems can be described to a satisfactory degree. Also, for model calibration, there is a significant lack of experimental data for a fine enough mesh of compositions for many alloys. While the by now ubiquitous high-throughput computations have provided data for many alloy properties, this is not true for metal-alloy conductivity since the electronic nature of heat transport there requires explicit treatment of electron scattering in alloys, which is to date mostly unexplored. The vast majority of predictive modeling of thermal conductivity has been performed for non-metals such as simple oxides and elemental semiconductors, where the thermal transport is dominated by lattice vibrations, which can be treated in a more straightforward way using molecular dynamics \cite{ref23}, perturbation theory \cite{ref24,ref25,ref26} and Monte-Carlo calculations \cite{ref27}.
	\bigbreak
	This lack of computational calibration data makes experimental data indispensible, which however due to the high cost and low efficiency of traditional measurement techniques, were previously scarce \cite{ref28}, resulting in an inability to develop and calibrate thermal conductivity models on the CALPHAD level. However, this situation has been changed through the recently proposed coupling of diffusion multiples with well-developed micron-scale spatial resolution properties microscopy tools (such as TDTR) \cite{ref29,ref44}. Diffusion multiples are blocks of three or more pure metal elements or alloys with a pre-designed geometry and intimate interfacial contacts to contain many diffusion couples and triples within one sample. After diffusion annealing at an elevated temperature, complete libraries of solid-solution phases and intermetallic compounds, with wide ranges of compositions, are formed. While the diffusion-multiple approach has long been used to determine phase diagrams and materials kinetics such as diffusion coefficients \cite{ref30,ref31,ref32}, gathering of high-throughput property data and building composition-structure-property relationship of alloys is a rather recent approach \cite{ref33}. This combinatorial approach, compared to the conventional properties measurements using individually made single-composition alloys, has orders of magnitude higher efficiency for study the composition-dependent properties, which we exploit here to develop for the first time a physical CALPHAD-level description of thermal conductivity in binary systems, accompanied by a full calibration for a series of binary noble- and transition metal alloys, specifically Au-Cu, Au-Ag, Pd-Ag, Pd-Cu, Pd-Pt, Pt-Rh, and Ni-Rh.
	\bigbreak
	The CALPHAD (CALculation of PHAse Diagram) approach \cite{ref1,ref2,ref3,ref4,ref5,ref6} was initially established as a tool for treatment of the composition- and temperature-dependent thermodynamic functions and phase equilibria from experimentally measured phase diagrams and thermochemical data. The merit of this approach is its ability to extrapolate from binary and ternary systems to multi-component systems to enable thermodynamic predictions of multicomponent phase diagrams and thermochemical properties for computational design of materials. The CALPHAD approach is also suitable for modeling phase-based properties using experimental data and physics-based models. This method has been successfully applied to diffusion coefficients in multicomponent systems \cite{ref7,ref8} and forms the basis for diffusion process simulations. A successful description of properties in CALPHAD requires both a solid physical model framework that allows interpolation and expansion to multi-component alloys, as well as sufficient composition-dependent validation and calibration data. Recent successful additions of alloy properties include molar volume, coefficient of thermal expansion (CTE) \cite{ref9} and elastic constants \cite{ref10}. With the resulting ability to predict the desired properties of multicomponent materials, the CALPHAD approach has been successfully used to design a number of new materials \cite{ref11,ref12}. Among the design-crucial properties that are currently not yet well included in the CALPHAD framework, thermal conductivity is arguably one of the leaders. Although semi-empirical models of thermal conductivity of insulators in which thermal conduction is dominated by phonons have been proposed by Gheribi and Chartrand \cite{ref13}, for metals and alloys in which the thermal transport is dominated by electrons, there are currently no reliable models available in the CALPHAD framework, which the current work will provide.
	
	\section{Theory}
	\subsection{Nordheim rule}
	
	For a completely soluble binary materials system (A-B solid-solution with $\textit{x}_A$ mole fraction of A and $\textit{x}_B$ mole fraction of B), the Nordheim rule \cite{ref20} proposes that the residual resistivity can be described by
	
	\begin{equation}\label{eq1}
	\rho(x_B)=Cx_Ax_B
	\end{equation}
	
	where $\rho$ is the residual resistivity and $C$ is a constant for all compositions.
	\bigbreak
	The Nordheim rule assumes random distributions of solute atoms in single phase solid solutions, without any phase mixtures. In addition, it is assumed that there is no significant change in crystal structure, atomic volume, and number of free electrons during alloying. The Nordheim rule has been successfully applied to a number of binary systems to describe the concentration-dependent  thermal conductivity or electrical resistivity, such as phonon dominated thermal conductivity of Si-Ge alloy \cite{ref34} and \textit{s-s} electron scattering dominated electrical resistivity of noble metal alloys \cite{ref35}. For alloys or systems containing transition metal elements, the Nordheim rule inadequately describes the electronic interactions, as detailed below.
	
	\subsection{Mott{\textquoteright}s two-band conductivity model}
	
	The localization of electronic states determines the nature of the associated electrons as itinerant, where they can travel through the crystal and carry charge or heat, or localized, where they are mostly immobile. The band structure, defined in reciprocal space and thus {\textquotedblleft}reciprocal{\textquotedblright} to real-space extent of the electronic states, gives a measure of the degree of localization: itinerant electrons have wide bands, while narrow bands indicate localized electrons. For alloys containing transition metal elements, the wide \textit{s}-valence bands house the conduction electrons traveling around, while the narrow \textit{d}-bands indicate localized electrons, which only scatter the \textit{s}-electrons, but contribute to conduction to a negligible degree.
	\bigbreak
	This division into itinerant and localized electrons was first proposed by Mott in the 1930{\textquoteright}s as a basis for resistivity modeling in alloys \cite{ref21,ref22}. While this model assumption is generally accepted and was used as a basis for the present discussion, we point out that positron-annihilation experiments have shown that \textit{d}-electrons can also be itinerant in a number of metals \cite{ref36}, suggesting a straightforward potential future expansion of our model where needed for such metals. 
	\bigbreak
	According to Mott{\textquoteright}s model, electrical resistivity is controlled by scattering of the itinerant \textit{s} electrons on impurities/solutes (or more exactly, their electrons) and phonons into vacant \textit{s}- and \textit{d}-states. Thus, the scattering rate is proportional to the density of states (DOS) of these vacant states at the Fermi level \cite{ref21}. He proposed to use Fermi{\textquoteright}s Golden Rule for the scattering probability $P(kk')$ per unit time for electrons transitioning from state $k$ to state $k'$,
	
	\begin{equation}\label{eq2}
	P(kk')= \langle\psi_{k'}|\Delta V|\psi_k\rangle^2g[E(k')]
	\end{equation}
	
	where  $\psi_k$ and $\psi_{k'}$ are the wave functions of the itinerant electrons before and after transition,  $\Delta V$ is the scattering potential introduced by the lattice distortion from alloying, and $g[E(k')]$ is the DOS of the final state $k'$.
	\bigbreak
	Mott carefully studied the Pd-Ag system as an example. He argued that in pure Ag, only \textit{s}-states were available at the Fermi level, and thus these dominated the DOS in Eq.~(2). When Pd was added into Ag, excessive \textit{s}-electrons from Ag filled the open \textit{d}-states of Pd up to about 40 at\% Pd composition. After that, there were insufficient \textit{s}-electrons from Ag to fill all of the vacant states in the \textit{d}-band of Pd, and \textit{d}-states became available as final scattering states. That was the reason that the electrical resistivity versus Pd concentration curve exhibited a sharp increase at that point. From this, Mott concluded that the total resistivity could be taken as the sum of \textit{s-s} and \textit{s-d} electron scattering. Intraband \textit{s-s} electron scattering followed the Nordheim rule and interband scattering from \textit{s-d} electrons hybridization was proportional to the \textit{d}-DOS of Pd in the alloy.
	\bigbreak
	In Mott{\textquoteright}s model, the \textit{s}-DOS is considered very small compared to the \textit{d}-DOS and thus ignored in the model. The \textit{d}-DOS of Pd is approximated as a simple parabolic function. With this, a satisfactory agreement was achieved between the model and the experimental data as shown in Figure 1(a). Mott{\textquoteright}s resistivity model thus consists of
	
	\begin{equation}\label{eq3}
	\left\{
	\begin{array}{c}
	\rho = \rho_{ss}+\rho_{sd}\\
	\rho_{ss}\propto x_{Pd}x_{Ag}\\
	\rho_{sd}\propto x_{Pd}x_{Ag}^2\cdot N_{d,Pd}(E_F)
	\end{array}
	\right.
	\end{equation}
	
	where $x_X$ is the mole fraction of element X in the alloy, and  $N_{d,Pd}(E_F)$ is the parabolic DOS of \textit{d}-electrons of Pd at the Fermi energy, 
	
	\begin{equation}\label{eq4}
	N_{d,Pd}(E_F)=\left\{\begin{array}{cc}
	0 & $if $ x> p \\
	(p-x)^2 & $if $ x\leq p
	\end{array}
	\right.
	\end{equation}

	where $p$ is a parameter describing how many open states are available per atom in the \textit{d}-band of pure Pd. In this model, Mott assumed $p = 0.65$ as indicated by the experimental magnetic susceptibility data from Svensson \cite{ref37}. However, as we can see from Figure 1(a), the fit with $p=0.65$ is suboptimal. If this parameter is set to be a free parameter, the best fit gives it a value close to 0.75.
	
	\bigbreak
	Coles and Taylor \cite{ref39} took the experimental electronic heat capacity data of several Pd-Ag alloys as the {\textquotedblleft}experimental{\textquotedblright} DOS of \textit{d}-electrons and used that in a model similar to the one of Mott. The agreement between their model and the experimental electrical resistivity is remarkable as shown in Figure 1(b), supporting Mott{\textquoteright}s approximation of the DOS shape and the subsequent resistivity model.

	\begin{wrapfigure}{r}{0.5\textwidth}\label{fig1}
		\centering
		\includegraphics[width=0.48\textwidth]{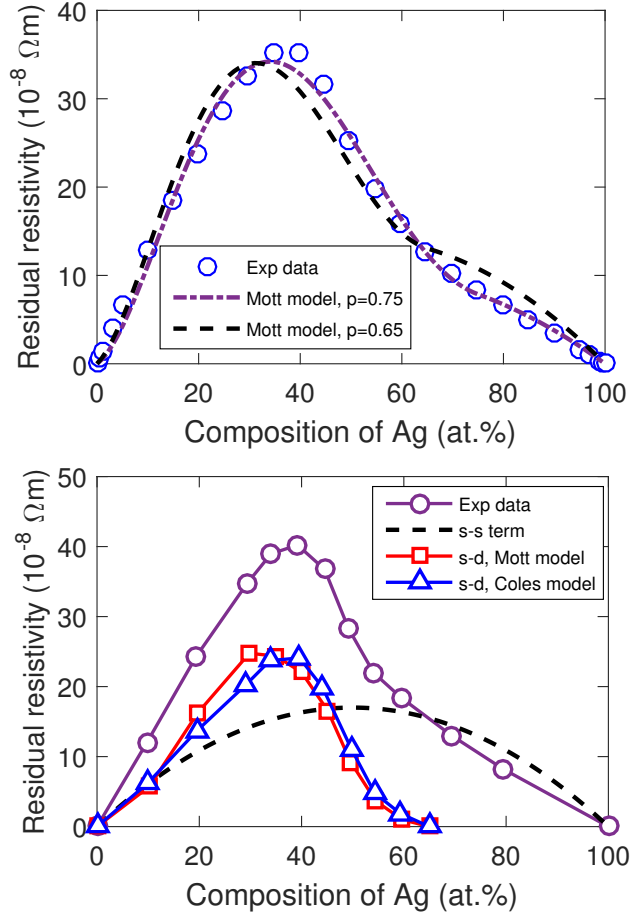}
		\caption{(top) Fit of Mott{\textquoteright}s model to the residual resistivity of Pd-Ag alloys. The black line is the model with $p=0.65$, the same value used in Mott{\textquoteright}s original paper. If $p$ is a free parameter, the best fit is achieved with $p = 0.75$ (magenta line). The experimental data are from Ref.~\cite{ref38}. (bottom) Coles and Taylor{\textquoteright}s model with electronic heat capacity data as DOS of \textit{d}-electrons. The magenta circles are the experimental residual resistivity data; the blue triangles are the \textit{s-d} resistivity after subtracting the \textit{s-s} part, $\rho_{sd}=\rho-Kx(1-x)$; the black dashed line is a fit of $Kx(1-x)$ by assuming $\rho_{sd}=0$ at $x=0.65$; the red squares are the \textit{s-d} resistivity as calculated with Mott{\textquoteright}s model with $x = 0.65$. Data in the figures were taken from Ref.~\cite{ref39}.}
	\end{wrapfigure}

	\bigbreak
	It is worth mentioning that even though Mott{\textquoteright}s two-band conductivity model agrees reasonably well with the experimental data, it still has a few flaws. Firstly, it assumes the DOS of \textit{d}-electrons of pure Ag is zero and thus it is not included in the model. This assumption is incorrect according to the \textit{ab initio} calculation results below. Moreover, for the transition metal-transition metal alloys, the \textit{s-d} scattering occurs at both alloy elements sites, and thus a single parabola cannot fit the total \textit{d}-DOS. Therefore both elements need to be included in the model. Secondly, the assumption that the DOS of \textit{d}-electrons at the Fermi energy is a parabolic function of alloy composition has limited applicability in real cases where much more complicated forms are found as shown in our results. Thirdly, the carrier concentration in the alloy is assumed to be independent of alloy composition in Mott{\textquoteright}s model. However, this is not consistent with first-principles results for the total DOS of the \textit{s}-electrons near the Fermi energy. In order to accurately describe the composition dependence of the thermal conductivity, the discussion needs to include separate functions for both alloy elements, accurately calculated DOS values, and composition dependent carrier densities.
	
	\subsection{Proposed general model --- \textquotedblleft Mott$^{+}$\textquotedblright}
	
	Based on Nordheim rule and Mott{\textquoteright}s two-band conductivity model, we propose a generalized model (Mott$^{+}$ model) for binary metal alloy systems, which can be extended to ternary or even multi-component systems if the partial densities of states were calculated by first-principles.
	\bigbreak
	For a solid solution alloy $A_{x_A} B_{x_B}$, in virtual crystal approximation, the lattice potential is described as
	
	\begin{equation}\label{eq5}
	V(x)=x_AV_A+x_BV_B.
	\end{equation}
	
	The resulting perturbation acting on the sites A can be calculated as
	
	\begin{equation}\label{eq6}
	dV_A=V_A-V(x)=x_B(V_A-V_B)=x_B\Delta V.
	\end{equation}
	
	Since $x_A+x_B = 1$, the total scattering probability is contributed from scattering occurring at A and B sites. Consistent with Mott{\textquoteright}s proposition, \textit{s}-electrons are the dominant conduction electrons due to their delocalization and much higher mobility; scattering between the conduction electrons and the open states in the \textit{d}-band and \textit{s}-band at the Fermi level is taken as the major source of the resistivity; and the probability of \textit{s-s} and \textit{s-d} scattering is proportional to the DOS of the final state at Fermi level \cite{ref40}.
	\bigbreak
	With these assumptions and Eq.~(6), the scattering probability from A atom sites becomes within Fermi{\textquoteright}s Golden Rule
	
	\begin{equation}\label{eq7}
	p_{A,s-s}=x_A\langle\psi_{s,k'}^A|dV_A|\psi_{s,k}^A\rangle^2g_s^A(E_F)=x_Ax_B^2\langle\psi_{s,k'}^A|\Delta V|\psi_{s,k}^A\rangle^2g_s^A(E_F),
	\end{equation}
	
	\begin{equation}\label{eq8}
	p_{A,s-d}=x_A\langle\psi_{d,k'}^A|dV_A|\psi_{s,k}^A\rangle^2g_d^A(E_F)=x_Ax_B^2\langle\psi_{d,k'}^A|\Delta V|\psi_{s,k}^A\rangle^2g_d^A(E_F),
	\end{equation}
	
	where  $\psi_{s,k}^A$ and $\psi_{s,k'}^A$ are the wave functions of \textit{s}-electrons of an A atom, before and after a scattering event, respectively. $\psi_{d,k}^A$ and $\psi_{d,k'}^A$ are the wave functions of \textit{d}-electrons, before and after a scattering event. $g_s^A(E_F)$ and $g_d^A(E_F)$ are the DOS of \textit{s}- and \textit{d}-electrons at the Fermi level, which are calculated from first-principles. Considering a scattering potential that is approximately constant for a given species at a given composition, the resulting integrals over the wave functions give DOS functions, and
	
	\begin{equation}\label{eq9}
	\langle\psi_{s,k'}^A|\Delta V|\psi_{s,k}^A\rangle^2\approx C_1'\Delta V_A^2\left|\langle\psi_{s,k'}^A|\psi_{s,k}^A\rangle\right|^2\approx C_1 V_A^2g_s^A(E_F)g_s^A(E_F),
	\end{equation}
	
	\begin{equation}\label{eq10}
	\langle\psi_{d,k'}^A|\Delta V|\psi_{s,k}^A\rangle^2\approx C_2'\Delta V_A^2\left|\langle\psi_{d,k'}^A|\psi_{s,k}^A\rangle\right|^2\approx C_2 V_A^2g_d^A(E_F)g_s^A(E_F).
	\end{equation}
	
	There, $V_A$ represents the (constant) scattering potential of an A atom, and $g_s^A (E_F)$ and $g_d^A (E_F)$ are again the DOS of \textit{s}- and \textit{d}-electrons at the Fermi level. The fitting coefficients $C$ are introduced here to account for differences between the exact value of the scattering interaction integral and our approximation using a constant scattering potential; this can be interpreted as a sort of \textquotedblleft efficiency factor\textquotedblright for the scattering process.
	\bigbreak
	Combining these terms with Eqs.~(7) and (8) and considering the scattering of all \textit{s}-electrons, we find the total scattering probability at A atom sites to be
	
	\begin{equation}\label{eq11}
	p_{A,s-s}=C_1 x_Ax_B^2 V_A^2g_s^A(E_F)^2g_s^{total}(E_F),
	\end{equation}
	
	\begin{equation}\label{eq12}
	p_{A,s-d}=C_2 x_Ax_B^2 V_A^2g_d^A(E_F)^2g_s^{total}(E_F).
	\end{equation}
	
	We consider the scattering potential $V_A$  to be species specific and proportional to the local average potential around atoms of a given species as calculated from first principles. We use the peak potential around each atomic site for the fitting results presented here, but find no significant difference in using an integrated average over the Bader volume for each atom.
	\bigbreak
	The ratio of the \textit{s-d} and \textit{s-s} scattering probability can be obtained according to Eqs.~(11) and (12) as
	
	\begin{equation}\label{eq13}
	\frac{p_{A,s-d}}{p_{A,s-s}}\approx \frac{g_d^A(E_F)^2}{g_s^A(E_F)^2}.
	\end{equation}
	
	We will evaluate this term in the context of real material systems in Section 5.
	\bigbreak
	Examining the expression for electrical conductivity, $\sigma=ne\mu$, we now see that our scattering probability should be proportional to $1/\mu$. We therefore need a term representing the carrier density in our material, the other factor that will vary with alloy composition. Assuming that the carriers contributing to conductivity are primarily unbound \textit{s}-electrons at the Fermi level, we can reuse the total \textit{s}-DOS from our \textit{ab-initio} calculations as the effective carrier density, such that our total Mott$^{+}$ resistivity as a function of composition $x_A$ will be
	
	\begin{equation}\label{eq14}
	\centering
	\begin{multlined}
	\rho(x_A)=x_A\rho_A+x_B\rho_B+\frac{1}{g_s^{total}(E_F)}\left(p_{A,s-s}+p_{A,s-d}+p_{B,s-s}+p_{B,s-d}\right) \\
	=x_A\rho_A+x_B\rho_B+C_1 x_Ax_B^2 V_A^2g_s^A(E_F)^2+C_2 x_Ax_B^2 V_A^2g_d^A(E_F)^2 \\
	+C_3 x_A^2x_B V_B^2g_s^B(E_F)^2+C_4 x_A^2x_B V_B^2g_d^B(E_F)^2.
	\end{multlined}
	\end{equation}
	
	Here $\rho(x_A)$ is the electrical or thermal resistivity of the alloy with a composition of $x_A$ mole fraction of A and $x_B$ mole fraction of B. The contribution $x_A \rho_A+x_B \rho_B$ to the total resistivity represents the intrinsic resistivity of the alloy resulting from the resistivity of the end compounds. $C_1$, $C_2$, $C_3$ and $C_4$ are the fitting coefficients. In the present work, this model is employed to fit composition dependent experimental (or literature) thermal resistivity data.
	
	\begin{wrapfigure}{L}{0.5\textwidth}\label{figmultiple}
		\centering
		\includegraphics[width=0.48\textwidth]{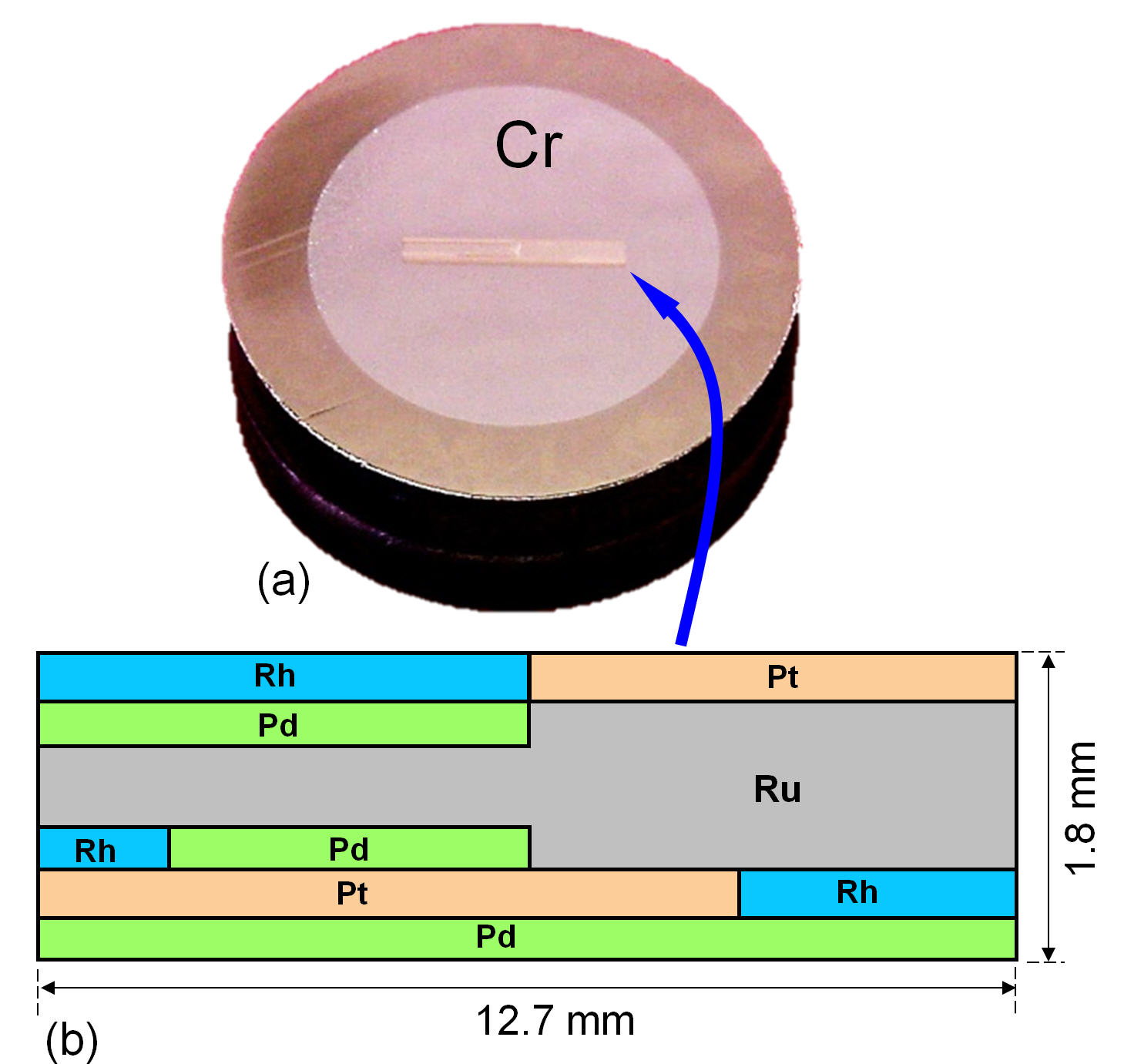}
		\caption{Cr-precious metal diffusion multiple for thermal conductivity measurement. (a) Optical photograph of the sample. (b) Layering inside the Cr piece.}
	\end{wrapfigure}
	
	\section{Experimental Procedures}
	
	The experimental data collected in this study were from high-throughput measurements on a Cr-based precious metal diffusion multiple Cr-Pd-Pt-Rh-Ru, shown in Figure 2. The sample was made by bricklaying pure Pd, Pt, and Rh foils, along with a piece of Ru, into a slot made in a Cr disk with a diameter of 25 mm and 3 mm in thickness. Two additional pure Cr disks (without the slot) with 25 mm in diameter and 3 mm in thickness were placed on the top and bottom of the slotted Cr disk. The whole assembly was placed in a commercially pure Ti can and sealed using electron beam welding in vacuum. Intimate surface contact among the metal pieces was achieved by hot isostatic pressing (HIP) at 1200 {\textdegree}C for 4 hours. After HIP, the sample was further annealed at 1200 {\textdegree}C for 36 hours in a quartz tube with backfilled argon atmosphere, and water quench to retain the high temperature phases. The diffusion multiple was cut into two halves by electrical discharge machining (EDM). Metallographic sample preparation was followed to ensure a good surface for further composition analysis and properties measurement. This diffusion multiple was originally used for mapping of phase diagrams, composition-dependent hardness and elastic modulus \cite{ref54}; key binary locations along the phase boundaries have been marked by nano-indentation patterns for later correlation of profiles of thermal conductivity and chemical composition.
	\bigbreak
	The measurements of concentration-dependent thermal conductivity of binary metallic alloys were made on the diffusion multiple using the time-domain thermoreflectance (TDTR) technique with a spatial resolution of $\sim3$ $\mu$m and a data acquisition speed of $\sim3$ points per second. TDTR is an optical pump-probe measurement technique that utilizes two pulsed laser beams split from a single beam \cite{ref42,ref43}. The pulse width is typically $\sim150$ femtoseconds. The modulated pump beam heats up the sample surface and the generated heat at the surface diffuses into the substrate via thermal conduction. The temperature excursion at the surface is measured by the intensity of the reflected probe beam through the change of the reflectivity as a function of temperature, i.e., thermoreflectance $dR/dT$. The arrival time of the probe beam is intentionally delayed compared with that of the pump beam by shortening the optical path of the pump beam using a 600 mm translational mechanical stage. Thermal properties of the sample can be determined by comparing the temporal evolution of the surface temperature with model calculations \cite{ref44}.
	\bigbreak
	TDTR has been demonstrated as a versatile and robust measurement technique of thermal conductivity \cite{ref45}, coefficient of thermal expansion (CTE) \cite{ref46}, and heat capacity ($\textnormal C_P$) \cite{ref47}. Benchmark studies show the accuracy of this technique on measuring thermal conductivity is within $±8$\%. Zheng et al.~\cite{ref46} have applied this method to measurement of thermal conductivity of several Ni based binary solid solutions and validated Wiedemann-Franz law. Zhao et al.~\cite{ref45} measured the concentration dependent thermal conductivity of the solid solutions and intermetallic compounds in the Ni-Al binary system using a Ni-NiAl diffusion couple. The comparison with literature values measured on individual alloys is excellent. In this paper, part of the thermal conductivity data of the binary systems used for the model validation were measured from the same experimental setup used in Ref.~\cite{ref47}, with the diffusion multiple arranged as pictured in Fig.~2.

	\begin{wrapfigure}{r}{0.5\textwidth}\label{fig2}
		\centering
		\includegraphics[width=.48\textwidth]{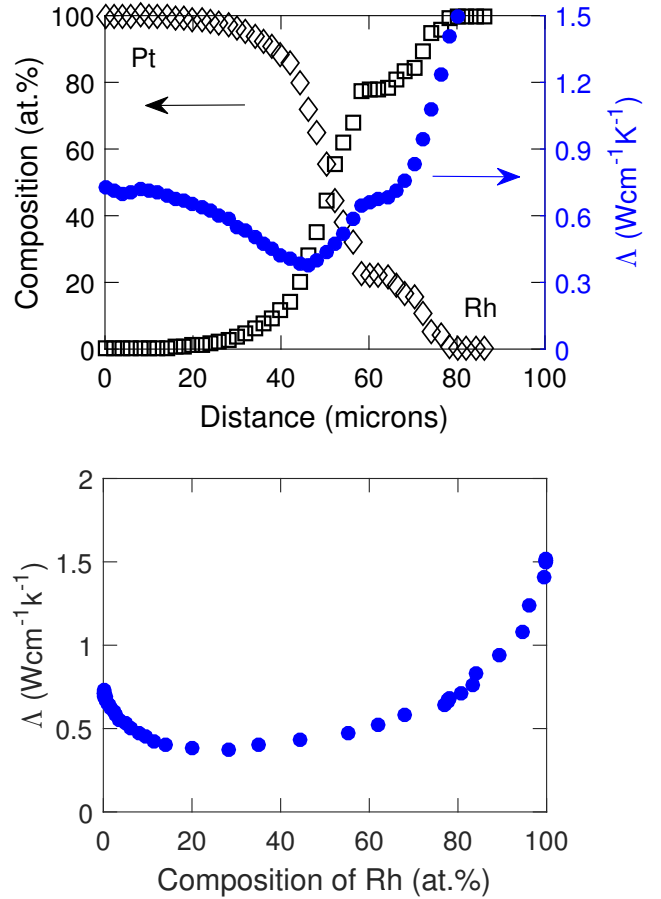}
		\caption{Composition profile from EPMA measurement and thermal conductivity profile from TDTR (top) were carefully aligned referenced to the dead centers of the indents to get the composition dependent thermal conductivity of Pt-Rh alloy (bottom).}
	\end{wrapfigure}
	
	\bigbreak
	The Pt-Rh binary system was used as an example to illustrate the process of how the concentration dependent thermal conductivity of each binary alloy system was obtained. Nano-indentation patterns were first marked on the sample surface at the pure-element ends of the diffusion zone, and the chemical composition profile between the indents was obtained through EPMA. The sample was then polished a little bit to get rid of the electron {\textquotedblleft}burn marks\textquotedblright which was usually introduced in the EPMA process and coated with $\sim100$ nm Al film using magnetron sputtering at room temperature. After coating, TDTR 2-D mapping was performed on the same area including the nanoindents to establish the thermal conductivity profile between the indents. The nano-indentation patterns marked on the sample surface act as a good reference to correlate the two profiles, thus we can accurately establish the compositional dependence of the thermal conductivity. The resulting chemical composition and thermal conductivity profiles are shown in Fig.~3(a), and the resulting concentration	dependent thermal conductivity in Fig.~3(b). Complete data sets from these experiments are included in the Supplementary Information.
	
	\section{Density of states (DOS) and scattering potential calculations}
	
	All calculations were performed using the Vienna Ab-initio Simulation Package (VASP) software package \cite{ref48,ref49} based on first-principles density functional theory. Generalized gradient approximation as given by Perdew-Burke-Ernzerhof (GGA-PBE)\cite{ref50} was used as the electronic exchange correlation energy in the main part of the calculation. Spin polarization was only considered in the case of systems that exhibit magnetic properties, i.e., Ni-Rh. Since the crystal structures for all the metallic systems of interest are face centered cubic (fcc) and the crystal structure is isomorphous for all compositions, the simulation cell is chosen to be fcc. Supercells composed of $4\times4\times4$ primitive cells with a total of 64 atoms were used with different ratios of atoms in solid solution to achieve the full composition range of each alloy system. A Monte Carlo code within the ATAT package \cite{ref51} was used to generate special quasirandom structures (SQS) of the simulation cell at each composition. A plane wave cut-off energy of 30\% higher than the standard cut-off energy was used along with a $6\times6\times6$ k-point mesh in all calculations. Complete data sets generated by these calculations are included in the Supplementary Information.
	
	\section{Results and Discussion}
	
	Non-linear fitting of the Mott$^{+}$ resistivity model from Eq.~(14) was performed with DOS data extracted from VASP calculations of the Au-Ag, Au-Cu, Ag-Pd, Cu-Pd, Pd-Pt, Pt-Rh, and Ni-Rh systems. Experimental resistivity data were used from both diffusion couple measurements in this study and reviews of resistivity measurements for transition metal systems \cite{ref41}.
	\bigbreak
	Previous derivations of theoretical alloy conductivity consider scattering from \textit{s}- and \textit{d}-electrons to be approximately equivalent, resulting in resistivity contributions that scale proportional to the relative DOS \cite{ref40}; our model does not make this assumption but instead allows the fitting parameter to adjust for the relative importance of all explicit orbital-resolved partial DOS contributions of all atomic species in the alloy system.
	
	\begin{wrapfigure}{r}{0.5\textwidth}\label{fig3}
		\centering
		\includegraphics[width=0.48\textwidth]{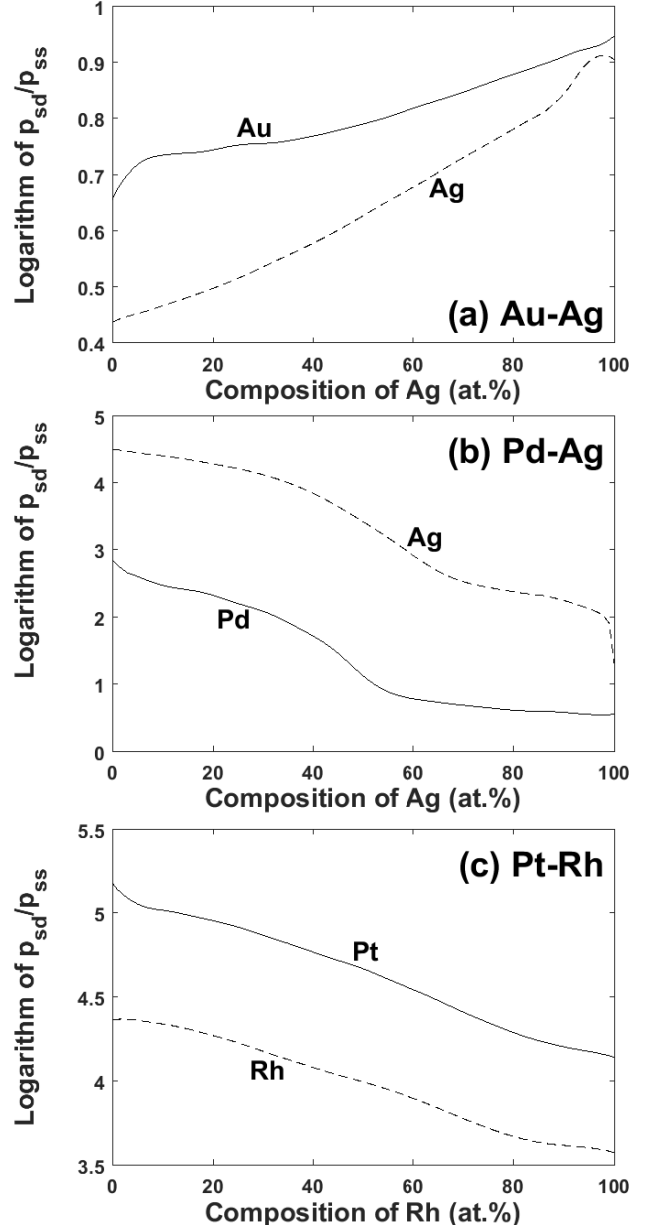}
		\caption{Relative scattering probability in the (a) Au-Ag, (b) Pd-Ag, and (c) Pt-Rh systems.}
	\end{wrapfigure}
	
	\bigbreak
	Prior to fitting, we consider the relative contribution of \textit{s-s} and \textit{s-d} scattering sites to the overall scattering probability in the alloy system as determined from Eq.~(13). Figure 4 shows three model cases: (a) the Au-Ag system, in which the scattering probability of \textit{s-s} sites is within a single order of magnitude of the scattering probability of \textit{s-d} sites across the entire composition range; (b) the Ag-Pd system, in which the scattering probability of \textit{s-s} sites is within a single order of magnitude of the scattering probability of \textit{s-d} sites for a portion of the composition range; and (c) the Pt-Rh system, in which the scattering probability of \textit{s-s} sites is smaller by three orders of magnitude than the scattering probability of \textit{s-d} sites across the entire composition range. It is possible in all cases to construct a non-linear fit of the experimental resistivity data using both \textit{s-s} and \textit{s-d} scattering terms; however, while the numerical quality of fitting is always increased by the inclusion of a greater number of fitting terms, the physical meaning of the fit can suffer via the inclusion of negative fitting coefficients when terms are included that are orders of magnitude smaller than the dominant contributing terms. This is unphysical since a negative fitting coefficient would imply that certain scattering sites reduce the frequency of carrier scattering within the alloy system. We find that these unphysical coefficients disappear when only terms with relatively large scattering potential over at least some portion of the composition range are included in the resistivity model.
	
	\bigbreak
	In order to evaluate the {\textquotedblleft}goodness\textquotedblright of the fit, we use a parameter called normalized sum of squared errors of prediction (SSE), which is defined as
	
	\begin{equation}\label{eq16}
	SSE=\Sigma_{i=1}^m\frac{(h_\theta(x^{(i)})-y^{(i)})^2}{h_\theta(x^{(i)})}
	\end{equation}
	
	where $h_\theta(x^{(i)})$ is the $i$-th element of the model calculation, $y^{(i)}$ is the $i$-th value of the experimental/literature data, and $m$ is the total data points. The fitting parameters and the SSE of the model fit are summarized in Table 1.
		
	\begin{figure}[H]\label{fig4}
		\begin{subfigure}{\textwidth}
			\centering
			\includegraphics[scale=.9]{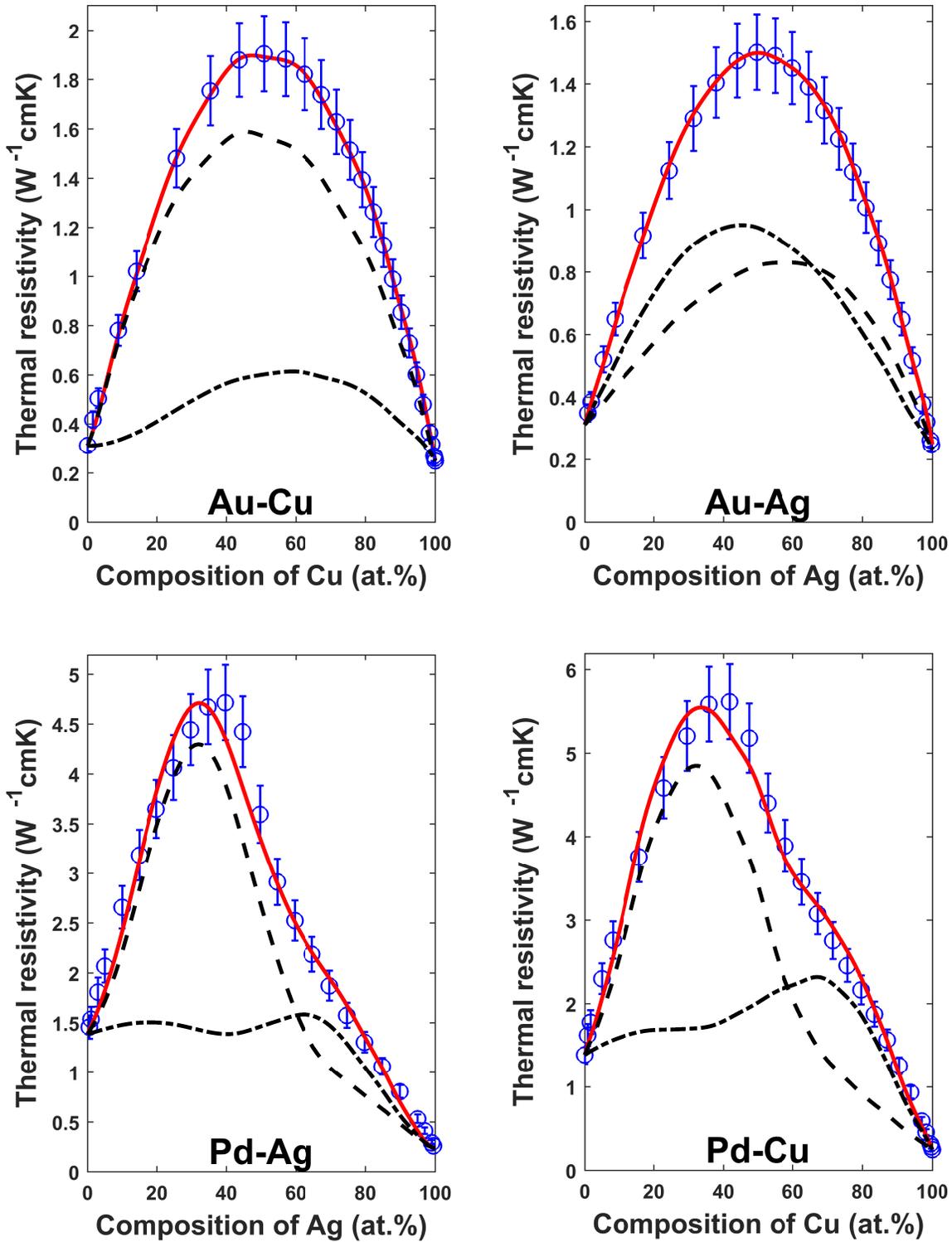}
		\end{subfigure}
		\caption{Mott$^+$ resistivity model fits to NIST recommended data from Refs.~\cite{ref38},\cite{ref41} for the Au-Cu, Au-Cu, Pd-Ag, and Pd-Cu systems and to experimental data for the Pd-Pt, Pt-Rh, and Ni-Rh systems. Error bars on NIST recommended data are displayed to be consistent with the experimental data collected in this study. (Continued on next page)}
		\end{figure}
		\begin{figure}[H]
		\ContinuedFloat
		\begin{subfigure}{\textwidth}
			\centering
			\includegraphics[scale=.9]{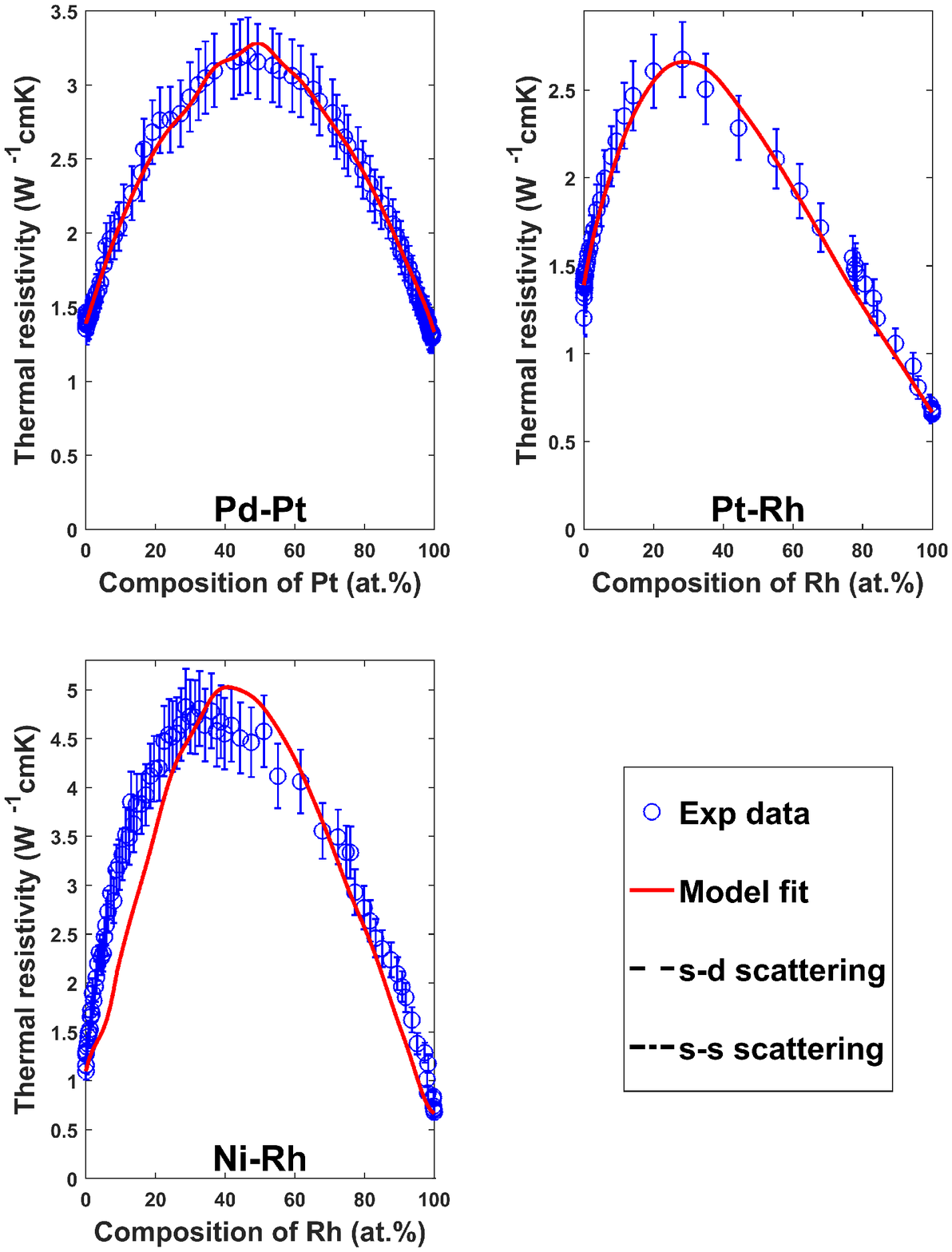}
		\end{subfigure}
		\caption{(Continued) Mott$^+$ resistivity model fits to NIST recommended data from Refs.~\cite{ref38},\cite{ref41} for the Au-Cu, Au-Cu, Pd-Ag, and Pd-Cu systems and to experimental data for the Pd-Pt, Pt-Rh, and Ni-Rh systems. Error bars on NIST recommended data are displayed to be consistent with the experimental data collected in this study.}
	\end{figure}
	\clearpage
	
	\begin{table}[ht]
		\centering	
		\caption{Resistivity model fitting data for the systems considered in this study. Fitting coefficients are in units of $\frac{\textnormal W^{-1} \textnormal{cm K}}{e^2 \textnormal{states per atom}}$}
		\small
		\begin{tabular}{|c|c|c|c|c|c|}
			\hline
			\textbf{System, A-B} & \textbf{\textit{s-s} coeff., A} & \textbf{\textit{s-d} coeff., A} & \textbf{\textit{s-s} coeff., B} & \textbf{\textit{s-d} coeff., B} & \textbf{SSE} \\
			 & \textbf{$(C_1)$} &  \textbf{$(C_2)$} & \textbf{$(C_3)$} & \textbf{$(C_4)$} & \\
			\hline
			Au-Cu & $1.70\times10^{-4}$ & $3.87\times10^{-5}$ & $6.95\times10^{-6}$ & $2.95\times10^{-5}$ & $4.66\times10^{-2}$ \\
			Au-Ag & $2.30\times10^{-4}$ & $4.95\times10^{-5}$ & $7.63\times10^{-5}$ & $8.56\times10^{-6}$ & $1.89\times10^{-2}$ \\
			Pd-Ag & $4.66\times10^{-4}$ & $3.16\times10^{-6}$ & $3.41\times10^{-3}$ & $1.63\times10^{-5}$ & $1.42\times10^{0}$ \\
			Pd-Cu & $4.49\times10^{-3}$ & $6.00\times10^{-6}$ & $4.05\times10^{-3}$ & $1.27\times10^{-6}$ & $1.09\times10^{0}$ \\
			Pd-Pt & - & $5.63\times10^{-7}$ & - & $1.53\times10^{-7}$ & $1.70\times10^{-1}$ \\
			Pt-Rh & - & $6.29\times10^{-7}$ & - & $1.95\times10^{-7}$ & $4.53\times10^{-1}$ \\
			Ni-Rh & - & $8.90\times10^{-7}$ & - & $2.64\times10^{-6}$ & $1.04\times10^{1}$ \\
			\hline
		\end{tabular}
		\normalsize
	\end{table}
	
	Figure 5 displays fits of our Mott$^{+}$ resistivity model to experimental data collected in this study as well as other sources \cite{ref41}. The uncertainty for these measurements is generally $±8$\% \cite{ref19}, as reflected by the error bars on the plots. For the Au-Ag, Au-Cu, Pd-Ag, and Pd-Cu systems, both \textit{s-s} and \textit{s-d} scattering terms are included in the plotted fits, while for the Pt-Pd, Pt-Rh, and Ni-Rh systems, only \textit{s-d} scattering terms are included in the plots, since for those systems, there was no composition range for which the \textit{s-s} scattering probability contributed significantly to the total carrier scattering.
	
	\bigbreak
	In the systems considered in this study, excellent agreement between the Mott$^{+}$ model and the experimental data is observed for all systems except the Ni-Rh system. This system has one notable difference that sets it apart from the other systems considered: it contains a magnetic phase. A full consideration of scattering in this system, then, would require consideration of scattering within each spin conduction pathway, as well as a consideration of spin mixing. Fitting of the Ni-Rh system was attempted in the formalism of Fert and Campbell \cite{ref55}, but it was found that without specific data as to the conduction pathways (far beyond the scope of this study), the fit is ill-conditioned and highly dependent upon initial conditions. It is anticipated that a full understanding of spin conduction within the end components as well as across the alloy compositions would be necessary to validate our model in a spin-polarized system.
	\bigbreak
	Equipped with our new understanding of alloy conductivity from first principles calculations, we may now go on to evaluate the effectiveness of previous empirical models at describing the physical process of electron scattering in an alloy system. The Nordheim rule, with its simplistic basis, is clearly only useful for symmetric empirical relationships between similar metals. The Mott model, on the other hand, contains sufficient parameters to consider the relationship between \textit{s-s} and \textit{s-d} scattering. We see that for most transition metal systems, \textit{s-d} scattering is by far the dominant term, as might be expected from the relative DOS around the Fermi level.  In fact, only in systems in which one or more of the constituents have completely filled \textit{d}-orbitals is the relative scattering potential of \textit{s-s} interactions great enough to make a meaningful contribution to the alloy resistivity, and even in those systems the resistivity is dominated by \textit{s-d} scattering contributions. With this consideration in mind, we evaluate Mott{\textquoteright}s two-band model as a reasonable approximation of alloy resistivity, based on inaccurate DOS data, resulting in its inaccurate assertion that significant contributions are seen from both the \textit{s-s} and \textit{s-d} scattering terms. We anticipate that a similar quality of fit could be achieved with any mathematical model consisting of the sum of two parabolas due to the shape of the DOS near the Fermi level across the composition range; this explains the relative agreement of the Mott two-band model even with its very primitive approximation of the DOS. The additional information our Mott$^{+}$ model provides is the relative importance of the scattering terms in each system: we can definitively say, based on our calculations, that \textit{s-d} scattering is the predominant scattering mechanism in all the transition metal systems considered in this work. We also note that our initial assumption that considering only electronic contributions to thermal conductivity would be sufficient for all transition metal systems becomes problematic in the case of systems in which thermal conduction can be influenced by spin-related processes, e.g., the Ni-Rh system considered in this work.
	
	\section{Conclusions}
	
	Composition dependent thermal conductivity data of several binary metal alloy systems were experimentally measured using time-domain thermoreflectance (TDTR). A generalized model was developed to correlate the thermal conductivity herein measured and previously reported with alloy composition, based on the DOS calculation of the \textit{s}- and \textit{d}-electrons near Fermi energy from first-principles. The overall agreement between the model and the experimental thermal resistivity data is excellent for all systems where electron transport and scattering are the dominant heat transport mechanism and allows to identify systems where additional mechanisms contribute significantly, such as spin-related scattering processes or magnon transport. If the related partial DOS at Fermi energy are calculated by first principles, this model can easily be applied to ternary or higher-order systems. This physical model is readily incorporated into the CALPHAD framework.
	
	\section*{Acknowledgements}
	
	The authors would like to thank Professor Joseph Heremans for insightful discussions and Professor Sheikh Akbar for his help in the Al film sputtering. C.W. and J.C.Z. were supported from the National Science Foundation under Grant No. DMR-0804833 and DMR-1237577. N.A. and W.W. acknowledge partial support from the Air Force Office of Scientific Research under Grant No. FA 9550-14-1-0322 and a DOE NEUP Fellowship. The calculations performed in this work were supported by an allocation of computing time from the Ohio Supercomputer Center \cite{ref53}.

\end{document}